\documentclass[%
 reprint,superscriptaddress,
 amsmath,amssymb,
 aps,prx
]{revtex4-2}
\usepackage{graphicx}
\usepackage{dcolumn}
\usepackage{bm}
\usepackage[caption=false]{subfig}
\usepackage{floatrow}
\usepackage{appendix}
\usepackage{amsmath}
\usepackage[italicdiff]{physics}
\usepackage[capitalise]{cleveref}
\usepackage[dvipsnames]{xcolor}

\DeclareCaptionLabelFormat{adja-page}{\hrulefill\\#1 #2 \emph{(previous page)}}
\DeclareRobustCommand{\l}{\left}
\DeclareRobustCommand{\r}{\right}

\begin{abstract}
Using determinant Quantum Monte Carlo, we compare three methods of evaluating the dc Hall coefficient $R_H$ of the Hubbard model:  the direct measurement of the off-diagonal current-current correlator $\chi_{xy}$
in a system coupled to a finite magnetic field (FF), $\chi_{xy}^{\text{FF}}$; the three-current linear response to an infinitesimal field as
measured in the zero-field (ZF) Hubbard Hamiltonian, $\chi_{xy}^{\text{ZF}}$; and the leading order of the recurrent expansion $R_H^{(0)}$ in terms of thermodynamic susceptibilities.  
The two quantities $\chi_{xy}^{\text{FF}}$  and $\chi_{xy}^{\text{ZF}}$  can be compared directly in imaginary time.  Proxies for $R_H$ constructed from the three-current correlator $\chi_{xy}^{\text{ZF}}$ can be determined under different simplifying assumptions and compared with $R_H^{(0)}$.  We find these different quantities to be consistent with one another, validating previous conclusions about the close correspondence between Fermi surface topology and the sign of $R_H$, even for strongly correlated systems.  These various quantities also provide a useful set of numerical tools for testing theoretical predictions about the full behavior of the Hall conductivity for strong correlations.
\end{abstract}

\begin{document}

\title{Numerical approaches for calculating the low-field dc Hall coefficient of the doped Hubbard model}

\author{Wen O. Wang}
\email{wenwang.physics@gmail.com}
\affiliation{Department of Applied Physics, Stanford University, Stanford, CA 94305, USA}

\author{Jixun K. Ding}
\affiliation{Department of Applied Physics, Stanford University, Stanford, CA 94305, USA}

\author{Brian Moritz}
\affiliation{Stanford Institute for Materials and Energy Sciences,
SLAC National Accelerator Laboratory, 2575 Sand Hill Road, Menlo Park, CA 94025, USA}

\author{Yoni Schattner}
\affiliation{Stanford Institute for Materials and Energy Sciences,
SLAC National Accelerator Laboratory, 2575 Sand Hill Road, Menlo Park, CA 94025, USA}
\affiliation{Department of Physics, Stanford University, Stanford, CA 94305, USA}

\author{Edwin W. Huang}
\affiliation{Department of Physics and Institute of Condensed Matter Theory, University of Illinois at Urbana-Champaign, Urbana, IL 61801, USA}

\author{Thomas P. Devereaux}
\email{tpd@stanford.edu}
\affiliation{Stanford Institute for Materials and Energy Sciences,
SLAC National Accelerator Laboratory, 2575 Sand Hill Road, Menlo Park, CA 94025, USA}
\affiliation{
Department of Materials Science and Engineering, Stanford University, Stanford, CA 94305, USA}
\date{\today}

\maketitle

\section{Introduction}
Transport measurements are among the most common and accessible experimental probes, and are often among the first to be performed following the discovery of new materials. Yet,
the theoretical investigation of normal state transport properties of 
{ quantum materials 
} presents a number of unique challenges.  
Fermi liquid theory and the associated Boltzmann transport theory, which provide the theoretical framework for the understanding of ordinary metals, are known to break down in certain regimes.  
In the case of the high-$T_c$ cuprates, this is evidenced by the linear-in-$T$ longitudinal resistivity, which violates the Mott-Ioffe-Regel (MIR) limit and has been synonymous with what has been called ``strange metallicity"~\cite{PhysRevLett.59.1337,PhysRevB.65.092405,PhysRevLett.93.267001,gunnarsson2003colloquium, hussey2004universality,calandra2003violation,gunnarsson2003colloquium, hussey2004universality,calandra2003violation,PhysRevLett.74.3253}.
In addition, the Hall coefficient $R_H$ for cuprates has a strong temperature dependence~\cite{PhysRevLett.84.3418,PhysRevB.75.024515}, in contrast to predictions of Fermi liquid theory.

The Hubbard model, despite its simple form, successfully captures some of the non-Fermi liquid signatures in the normal state of cuprates. 
 Strange metallic resistivity, without a signature of saturation at the MIR limit, has been successfully observed in the Hubbard model, both numerically~\cite{huang}  via determinant quantum Monte Carlo (DQMC)~\cite{DQMC1,DQMC2} simulations and in cold atoms experiments~\cite{xu2019bad,brown2019bad}.
 $R_H$  investigated in another recent DQMC work~\cite{wen} also shows  strong temperature dependence and a non-trivial peak at temperature $T \sim t$, the kinetic energy scale, which may be connected to the rise of $R_H$ in cuprates like LSCO as temperature decreases from the ultra high temperature limit~\cite{PhysRevLett.72.2636}.
The close relationship between the experimental observations of cuprates and the theoretical results from the Hubbard model motivate us to continue investigating transport properties of the Hubbard model, specifically $R_H$.  
We seek to better understand novel transport phenomena in materials without quasiparticles, 
and the relationship to the evolution of the electronic structure for understanding intertwined phases in the cuprates~\cite{keimer2015quantum}.

A calculation of the Hall coefficient is more complex compared to the longitudinal resistivity. 
Previous works have investigated $R_H$ in the $t-J$ and Hubbard models for both high frequency and dc limits~\cite{PhysRevB.66.020408,PhysRevLett.70.2004,stanescu2003full,PhysRevLett.74.3868,prelovvsek2001reactive,PhysRevB.59.1800,PhysRevB.4.1566}; however,
much of the work for strongly correlated models has involved certain simplifying assumptions, approximations, and limiting cases~\cite{assa, prelovvsek2001reactive}. A faithful comparison of methods that can be used to calculate the dc Hall coefficient in a strongly correlated framework has been lacking.

Measuring $R_H$ in a numerically exact way poses challenges, limited by both the speed and efficiency of numerical techniques.
For example, exact diagonalization is restricted to small lattice sizes~\cite{PhysRevB.49.5065}; 
density-matrix renormalization group (DMRG) is limited to a small number of exited states and may not be stable for calculating transport properties of 2D systems, especially in the metallic phase~\cite{jeckelmann2008dynamical}.
Numerical simulations for larger lattice sizes can be performed using quantum Monte Carlo (QMC) simulations in imaginary time for temperatures where the fermion sign problem is not too severe~\cite{SIGN}.  In the work presented here we use DQMC, a particular flavor of QMC.

One approach for obtaining $R_H=-B^{-1}\rho_{xy}$ via QMC simulations is to explicitly couple the Hubbard model to a finite magnetic field. Current-current correlators $\chi_{\alpha\beta}(\tau)$ ($\alpha$,$\beta = x$ or $y$ direction) measured in imaginary time are then analytically continued to real frequency to obtain all components of the conductivity tensor $\sigma_{\alpha\beta}(\omega)$.
In this approach, explicitly adding a magnetic field $B$ raises the computational complexity by requiring complex (as opposed to real) calculations.  Apart from the inherent difficulty in properly incorporating the magnetic field $B$ due to considerations of gauge invariance, this procedure also suffers from the need to analytically continue both the diagonal and off-diagonal components of $\sigma_{\alpha\beta}$ concurrently~\cite{ana1, ana2}.

In an alternative approach, one could consider the zero-field limit by expanding the off-diagonal part of $\chi_{\alpha\beta}$ up to linear terms in $B$. This method still requires analytic continuation, but avoids measurements in a finite field. 
However, in this approach one must evaluate a correlation function of higher order fermion operators (six fermions in the Hubbard model), which can increase error propagation. In addition, by introducing an extra imaginary time and an extra space index, the simulation becomes computationally more expensive. We provide additional detail in Appendix~\ref{sec:simulationdetails}.

In Ref.~\cite{assa}, another route was laid for studying $R_H$ numerically. 
As an application of the recursion method~\cite{recursion1,recursion2},
this technique expands the Kubo formula of dc Hall conductivity in a Liouvillian representation into terms determined by magnetization matrix elements and Liouvillian matrix elements (or recurrents)~\cite{liou} in a Krylov basis.
By expanding $R_H=\sum_k R_H^{(k)}$, where  $R_H^{(k)}$ consist of thermodynamic susceptibilities,
the expansion avoids the need for analytic continuation~\cite{assa,assa2}. 
We refer to this method as the recurrent expansion. One drawback of this method is that the expansion is only conditionally convergent and its truncation error can be hard to estimate for strongly interacting systems.

In previous work~\cite{wen}, we investigated the dc Hall coefficient $R_H$ of the Hubbard model using 
DQMC to evaluate
the leading order of the recurrent expansion $R_H^{(0)}$, showing a strong temperature dependence - increasing with decreasing temperature - mimicking the behavior seen in cuprates~\cite{PhysRevLett.72.2636}. Despite the strong temperature dependence deviating from Fermi liquid behavior, the sign of $R_H$ displayed 
a surprisingly close relationship with the Fermi surface topology, which has been usually understood as a feature of free electrons.
As the interaction increases or the doping decreases towards half filling, $R_H$ changes sign concomitant to changes in Fermi surface topology.
We argued that the ``Hall coefficient sign -- Fermi surface topology" correspondence may apply even for very strong interactions and low doping, in close proximity to a Mott insulator.

Higher-order corrections in the recurrent expansion of $R_H$~\cite{assa,assa2} could be large enough to qualitatively change this behavior. 
In the expansion described in Refs.~\cite{assa,assa2}, the higher order $k^{\text{th}}$ magnetization matrix elements and recurrents are
constructed from correlators containing operators proportional to
$\dd^{k}J_{\alpha}/{\dd t^{k}}$,
where 
$J_\alpha$ is the current operator along the $\alpha = x$ or $y$ direction.
$\dd^{k}J_{\alpha}/{\dd t^{k}}$ may produce terms that include a number of fermion operators, which makes these higher order recurrents more computationally expensive to measure in comparison to $R_H^{(0)}$.

Since the convergence rate of the recurrent expansion is hard to determine away from weak coupling, and higher order corrections are expensive to calculate, we consider the two approaches mentioned previously, which  focus directly on the field response of $\chi_{\alpha\beta}$. 
As exact expressions measured in a well-controlled algorithm,  they can be compared with our result for $R_H^{(0)}$~\cite{wen}. The imaginary time dependence of $\chi_{\alpha\beta}$ also contains real-time dynamic information about the conductivities. 

In this work, we use numerically exact DQMC simulations to evaluate the dc Hall coefficient $R_H$ in the weak-field limit using multiple methods: 
\begin{itemize}
\item Recurrent Expansion -- leading order $R_H^{(0)}$ in terms of thermodynamic susceptibilities~\cite{wen,assa,assa2},  
\item Zero Field (ZF) -- the three-current linear-response of the off-diagonal part of the correlator 
{$\chi_{\alpha\beta}$ 
to first order in the magnetic field},  $\chi_{\alpha\beta}^{\text{ZF}}$.
\item Finite Field (FF) -- directly evaluating 
{  $\chi_{\alpha\beta}$  
} for a gauge invariant Hamiltonian in weak finite-fields on a finite-size lattice~\cite{katherine}, $\chi_{\alpha\beta}^{\text{FF}}$.
\end{itemize}
We compare results from the latter two methods directly in imaginary time, finding a high degree of consistency, demonstrating that the DQMC algorithm is well-equipped to handle orbital effects of magnetic fields.  To avoid the caveats of analytic continuation, we estimate various proxies for $R_H$ from the three-current correlation function. We find reasonable consistency with previous results for $R_H^{(0)}$~\cite{wen}. These findings reaffirm the correspondence between the sign structure exhibited by $R_H^{(0)}$ and the topology of the underlying Fermi surface, even in the limit of strong correlations that lack well-formed quasiparticles. 
In addition, we find that $R_H$ varies more slowly in Matsubara frequency than the individual longitudinal or transverse conductivities. We speculate that the cancellation of strong Matsubara frequency dependence of the individual conductivities also may be related to the observed correspondence between $R_H$ and the Fermi surface topology \cite{wen}.

The remainder of this paper is organized as follows. In section~\ref{3currderivations}, we discuss the inclusion of orbital magnetic fields into the DQMC algorithm, and provide an expression of the zero-field linear response $\chi_{xy}^{\text{ZF}}$ 
and show the comparisons between the ZF and FF results in imaginary time. In section~\ref{proxysec} we construct proxies for estimating the Hall coefficient from  $\chi_{xy}^{\text{ZF}}$ and   $\chi_{xx}$ (taken as the ZF longitudinal response)  and discuss the comparisons between them. We close with a discussion of our results and the challenges that remain for an evaluation of the full frequency dependence of the conductivities in the Hubbard model in a magnetic field.

\section{Current-current correlation functions in the presence of a magnetic field} \label{3currderivations}
In this section, we first discuss the inclusion of magnetic fields into the Hubbard model and derive an expression for the off-diagonal component of the current-current correlation function, to linear order in the magnetic field.  We compare this directly to the current-current correlation measured under the lowest nonzero allowed field in imaginary time.

Here and throughout the paper, we have neglected Zeeman coupling of applied magnetic fields to spins and focus solely on the orbital contributions relevant for the Hall 
conductivity. 
The Hamiltonian of the Hubbard model in the presence of an orbital magnetic field is
\begin{align}
\mathit{H}(B) = &-\mathit{t}\sum_{\langle \mathit{\mathbf{r_1},\mathbf{r_2}}\rangle
,\sigma} \mathit{c_{\mathbf{r_1},\sigma}^{\dagger} c_{\mathbf{r_2},\sigma}}
 \mathrm{e}^{\mathit{i}
 \mathit{\theta}_{\mathbf{r_1},\mathbf{r_2}}
 }
- \mu\sum_{\mathit{\mathbf{r}},\mathit{\sigma}} \mathit{n}_{\mathbf{r},\mathit{\sigma}} \nonumber \\
&+ \mathit{U}\sum_{\mathbf{r}}\mathit{n}_{\mathbf{r},\uparrow
} \mathit{n}_{\mathbf{r},\downarrow}
\stepcounter{equation}\tag{\theequation}\label{eq:hubbard},
\end{align}
where $\mathit{t}$ is nearest-neighbor hopping energy,
$\mathit{\mu}$ is the chemical potential, $\mathit{U}$ is the on-site repulsive interaction, $\mathit{c}_{\mathbf{r},\mathit{\sigma}}^{\dagger}$ $(\mathit{c}_{\mathbf{r},\mathit{\sigma}})$ is the creation (annihilation) operator for an electron at position $\mathbf{r}$ with spin $\mathit{\sigma}$, and $\mathit{n}_{\mathbf{r},\mathit{\sigma}} \equiv \mathit{c}_{\mathbf{r},\mathit{\sigma}}^{\dagger} \mathit{c}_{\mathbf{r},\mathit{\sigma}}$ is the number operator, with real-space lattice position $\mathbf{r}$ given by
$\mathbf{r} = x \mathbf{e}_x + y \mathbf{e}_y$, where $\mathbf{e}_x$ and $\mathbf{e}_y$ are unit vectors and the lattice constant is set to $1$. The model is placed on a square lattice 
and we use periodic boundary conditions such that $c_{r + L_x \mathbf{e}_x} \equiv c_{r + L_y \mathbf{e}_y} \equiv c_r$, unless otherwise specified, where $L_x$ and $L_y$ are the linear size of the system in the $x$ and $y$ directions, respectively. 
Here, 
$\mathit{\theta}_{\mathbf{r_1},\mathbf{r_2}} = \int_\mathbf{r_1}^\mathbf{r_2} \mathit{e}\mathbf{A}(\mathbf{r})\cdot d\mathbf{r}$ is the Peierls phase and $\mathbf A$ is the vector potential, and this Peierls phase integral is calculated using the shortest straight line path. Finally, the current density operator~\cite{mahan2013many} is given by 
\begin{equation}
J_\alpha(\mathbf{r};B) = 
-\mathit{i}\mathit{et} \sum_\sigma  c_{\mathbf{r}+\mathbf{e}_\alpha,\sigma}^{\dagger} c_{\mathbf{r},\sigma}
 \mathrm{e}^{\mathit{i}
 \theta_{\mathbf{r}+\mathbf{e}_\alpha,\mathbf{r}}}
+\mathrm{h.c.}. \label{eq:current}
\end{equation}
To circumvent the problem that the magnitude of a uniform field is limited to integer multiples of the flux quantum $\Phi_0 = 2\pi/e$ on a torus, we take the magnetic field to have a finite wavevector $B(\mathbf{r})= B \cos(q y)$ with $q=2\pi/L_y$ in finding the zero-field linear response expression.
With this choice, the field magnitude $B$ can be arbitrarily small while maintaining periodic boundary conditions. We have verified that for the systems sizes investigated here, the value of $q$ does not affect the overall results (See Appendix~\ref{sec:finitesize}).

Expanding the current-current correlation function to first order in $B$ yields,
\begin{align}
    &\chi_{xy}(\mathbf r,\tau) \equiv -\sum_{\mathbf{r}''} \left\langle T_\tau J_x(\mathbf{r}, \tau;B) J_y(\mathbf{r}'',0;B)\right\rangle_B /B \nonumber\\
    &=-\int_0^\beta \dd\tau' \sum_{\mathbf{r}' \mathbf{r}''} A_x(\mathbf{r}')\left\langle T_\tau J_x(\mathbf{r}, \tau) J_x(\mathbf{r}',\tau') J_y(\mathbf{r}'',0)\right\rangle/B + O(B), \label{first0}
\end{align}
where $T_\tau$ is the imaginary time ordering operator, $\beta=1/k_B T$, $T$ is the temperature, and we use the gauge where $\mathbf{A} (\mathbf{r}) =-\left(B \sin(q y)/q\right) \,\mathbf{e}_x$. Here, $\langle \cdots \rangle_B$ denotes the expectation value taken with the full Hamiltonian, while $\langle \cdots \rangle$ is taken with the $B=0$ Hamiltonian.~\footnote{Note that the translational and reflection symmetries of the unperturbed Hamiltonian imply that the modification of the current operator by the magnetic field does not contribute to $\chi_{xy}(\mathbf{r},\tau)$ to first order in $B$.}

Next, by making use of translation and reflection symmetries of the unperturbed Hamiltonian, we find
\begin{equation}
\chi_{xy}(\mathbf r, \tau) = B(\mathbf r) \chi^{\text{ZF}}_{xy}/B + O(B),\label{first00}
\end{equation} 
where 
\begin{align}
    \chi^{{\text{ZF}}}_{xy}(\tau) =
    &\int_0^\beta \dd\tau'\sum_{\mathbf{r}'\mathbf{r}''}\frac{\sin(q(y'-y''))}{q} \nonumber\\
    &\qquad\times\langle T_{\tau} J_{x}(\mathbf{r}'',\tau)J_{x}(\mathbf{r}',\tau')J_{y}(\mathbf{0},0)\rangle \label{eq:3current}
\end{align}
is the zero-field linear response, 
which we evaluate using DQMC simulations. 
We note in passing that a similar expression was derived for the case of a continuum model in the limit $q\rightarrow 0$~\cite{itoh1984exact,itoh1985gauge,fukuyama1969theory}.

In addition to $\chi^{\text{ZF}}_{xy}$, we also consider the current-current correlation function in a uniform magnetic field. As discussed above, the smallest uniform magnetic field that can be applied corresponds to a single flux quantum through the system $B=\Phi_0/V$, where $V=L_x L_y$ is the area of the system.
We use the gauge
$\mathbf{A}= B(-  y \mathbf{e}_x+   x \mathbf{e}_y)/2$
and the 
corresponding modified periodic boundary conditions
\begin{align}
c_{\mathbf{r}+L_x\mathbf{e}_x}  \equiv c_{\mathbf{r}}e^{-\mathit{i}\, eBL_x y /2}, \nonumber\\
    c_{\mathbf{r}+L_y\mathbf{e}_y}  \equiv c_{\mathbf{r}}e^{\mathit{i}\, eBL_y x /2}, \nonumber
\end{align}
following Ref.~\cite{PhysRevB.65.115104}. We then define the finite-field current-current correlation function as
\begin{equation}
    \chi^{\mathrm{FF}}_{xy}(\tau) = \frac{-1}{V B}\sum_{\mathbf{r},\mathbf{r}'} \left\langle T_\tau J_x(\mathbf{r},\tau;B) J_y(\mathbf{r}',0;B)\right\rangle_B. \label{FF}
\end{equation}

We evaluate $\chi^{\mathrm{FF}}_{xy}$ by performing a separate set of simulations in the presence of a magnetic field, and compare with $\chi^{{\text{ZF}}}_{xy}$. The two expressions are expected to agree in the thermodynamic limit.
Technical details of simulations can be found in Appendix~\ref{sec:simulationdetails}.

\begin{figure}[htbp]
    \includegraphics[width=\textwidth]{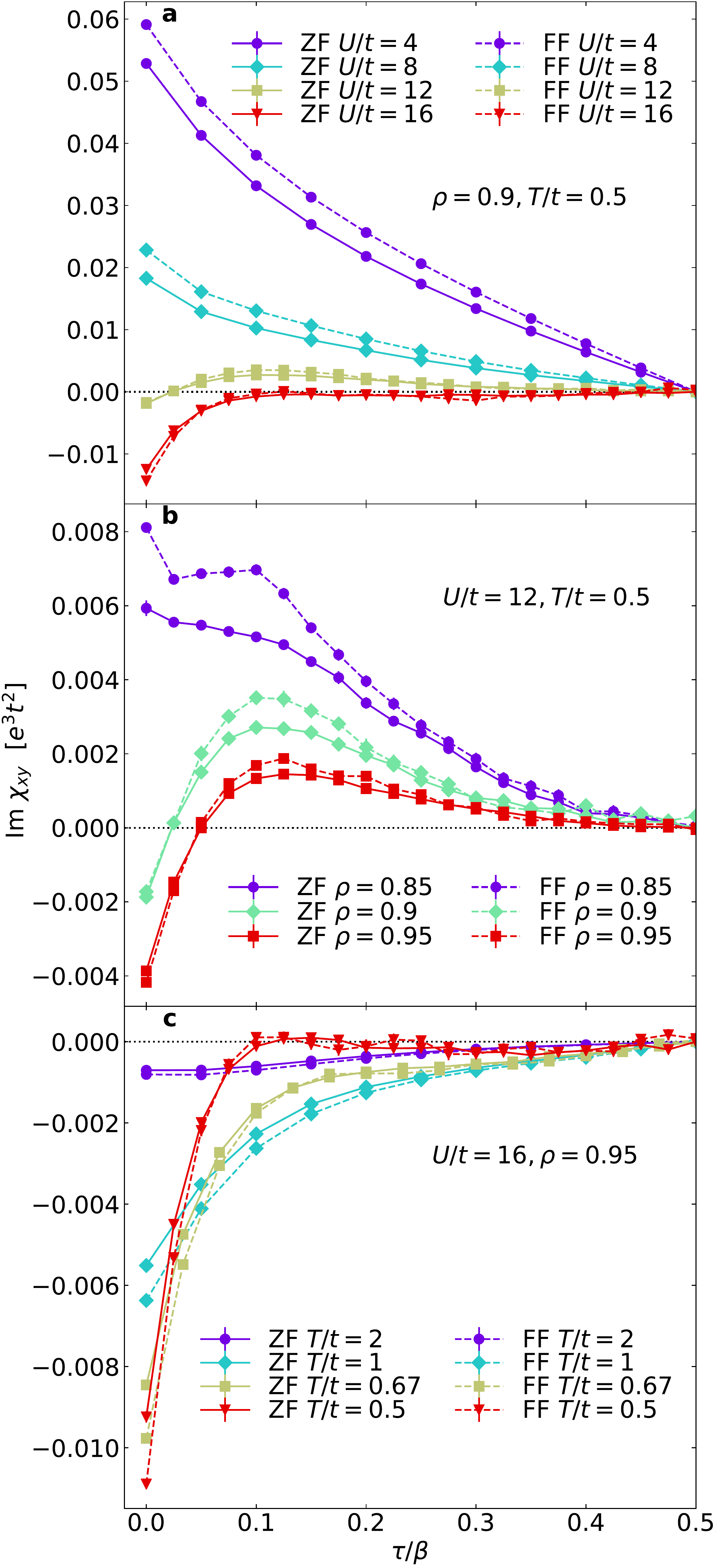} 
    \caption{ \label{fig:compare_finitefield}
    Comparisons between zero-field (ZF, solid lines) $\Im \chi_{xy}^{\text{ZF}}$ (Eq.~\ref{first0}) and finite-field (FF, dashed lines) $\Im \chi_{xy}^{\text{FF}}(\tau)$ (Eq.~\ref{FF}) in imaginary time for calculations on a $6\times 6$ lattice. 
    \textbf{a} A comparison at fixed filling ($\rho=0.9$) and temperature 
    ($T/t = 0.5$) for interaction strengths $U/t=4 - 16$. 
    \textbf{b} A comparison for fixed $U/t=12$ and  $T/t = 0.5$ for fillings $\rho=0.85 -0.95$.
   \textbf{c} A comparison for fixed $U/t=16$ and $\rho=0.95$ for temperatures 
   $T/t = 0.5 - 2$.
   The ZF and FF values have the same units -- $e^3 t^2$.  Error bars represent $\pm 1$ standard error of the mean, determined by jackknife resampling~\cite{jackknife}. 
    }
\end{figure} 

  While $\hbar=1$ for convenience, in natural units the unit of conductivity $\sigma$ is $e^2/\hbar$ and the unit of magnetic field strength
$B$ is $\hbar/(e a^2)=\hbar /e$ for lattice constant $a=1$. Therefore, our unit for $R_H$ is  $e^{-1}$.

Results for $\chi_{xy}^{\text{ZF}}(\tau)$ and $\chi_{xy}^{\text{FF}}(\tau)$ (both purely imaginary) plotted against imaginary time are shown in Fig.~\ref{fig:compare_finitefield}.
Generally, the transverse conductivity is reduced for large $U$, Fig.~\ref{fig:compare_finitefield}\textbf{a}, and as half filling is approached, Fig.~\ref{fig:compare_finitefield}\textbf{b}, as expected when charge fluctuations are suppressed at large $U$ and particle-hole symmetry is restored at half-filling. 
The zero field (ZF) result $\chi_{xy}^{\text{ZF}}(\tau)$ also qualitatively matches the finite field (FF) result $\chi_{xy}^{\text{FF}}$.
The significant features, including doping dependence, temperature dependence, and imaginary time dependence, agree quite well between ZF and FF, which implies that when converted to $\sigma_{xy}(\omega)/B$, the two methods should also produce similar frequency dependent features.
We verified that the small discrepancies between ZF and FF results are reduced for larger lattice sizes, as shown in Fig.~\ref{fig:finitesizeU8} of Appendix C. We henceforth suppress the labels FF and ZF from $\chi_{xy}^{\text{ZF/FF}}$ unless it is needed.

We observe that $\partial_\tau \Im \chi_{xy}$ at $\tau = \beta/2$ 
tends to increase with increasing $U$ or decreasing doping,
and may even change sign for certain parameters (see Fig.~\ref{fig:morefffzcompare} in the Appendix for more details).
As the off-diagonal conductivity is related to $\chi_{xy}$ 
by (Appendix~\ref{sec:conductivity})
\begin{align}
     &\chi_{x y}(\tau) = i\int_0^{\infty} \frac{\dd \omega}{\pi} \frac{\omega^2 \sinh\left[\omega(\tau-\beta/2)\right]}{\sinh(\omega\beta/2)} \frac{\mathrm{Im}\,\sigma_{xy}(\omega)}{\omega B}\nonumber \\
     &\equiv i\int_0^{\infty} \dd \omega \, K(\tau,\omega) \, \frac{\mathrm{Im}\,\sigma_{xy}(\omega)}{\omega B}, \label{Imrelation}
\end{align}
by symmetry, $\chi_{xy}(\tau = \beta/2) = 0$. For $U=0$, in the thermodynamic limit, $\partial_\tau \chi_{xy}$ does not depend on $\tau$. 
In this case, Eq.~\ref{Imrelation} leads to $\omega\,\text{Im}\, \sigma_{xy}(\omega)\propto \delta(\omega)$, as expected for infinite lifetime quasiparticles in the non-interacting limit. 
In Fig.~\ref{fig:compare_finitefield}\textbf{a}-\textbf{b}, we see that $\chi_{xy}$ measured under the weakest interaction strength ($U/t=4$) and lowest filling ($\rho=0.85$) shows similar $\tau$ behavior to that expected for $U=0$, and shows strong deviations with increasing $U$ or as the system approaches a Mott insulator at half filling.  The large curvature of $\Im \chi_{xy}(\tau)$ at small $\tau$ reflects the features of $\sigma_{xy}(\omega)$ at $\omega \sim U$ due to transitions to the upper Hubbard band. This becomes more pronounced at low temperatures and higher $U$ as evident in Fig.\ref{fig:compare_finitefield}\textbf{c}.

\section{Proxies}\label{proxysec}
We wish to obtain the zero frequency limit of the transverse conductivity from the imaginary time result shown in Fig.~\ref{fig:compare_finitefield}. Normally, this procedure for the longitudinal conductivity $\sigma_{xx}$ involves inverting 
\begin{align}
    \chi_{x x}(\tau) &= -\int_0^{\infty}\frac{\dd \omega}{\pi} \frac{\omega \cosh\left[\omega(\tau-\beta/2)\right]}{\sinh(\omega\beta/2)}  \Re \sigma_{xx}(\omega), \label{Rerelation}
\end{align}
where 
$ \chi_{xx}(\tau) \equiv -\sum_{\mathbf{r}',\mathbf{r'}}\ev{J_x(\mathbf{r},\tau),J_x(\mathbf{r}',0)}/V$. 
Since $\Re \sigma_{xx}(\omega)$ is positive definite, maximum entropy techniques (MEM)~\cite{ana1,ana2} may be employed to obtain $\sigma_{xx}(\omega)$ from $\chi_{xx}(\tau)$. However this is not the case for the imaginary part of the transverse frequency dependent conductivity $\Im \sigma_{xy}$, and as a result MEM techniques encounter difficulties. This can be seen in Fig.~\ref{fig:compare_finitefield} wherein $\Im \chi_{xy}(\tau)$ can change sign in the range $[0,\beta/2)$, while the kernel $K(\tau,\omega)$ in Eq.~\ref{Imrelation} does not change sign. 

In addition, the six-fermion correlator in $\chi_{xy}^{\text{ZF}}$ is computationally expensive to measure and suffers from large numerical errors. 
Therefore, in this section, to compare our
$\chi_{xy}^{\text{ZF}}$ result obtained from Eq.~\ref{eq:3current}, with $R_H^{(0)}$ obtained in Ref.~\cite{wen}, we construct proxies for dc $R_H$ using $\chi_{xy}^{\text{ZF}}$, and compare these proxies to $R_H^{(0)}$. 

\begin{figure*}
    \centering
    \includegraphics[width=0.9\textwidth]{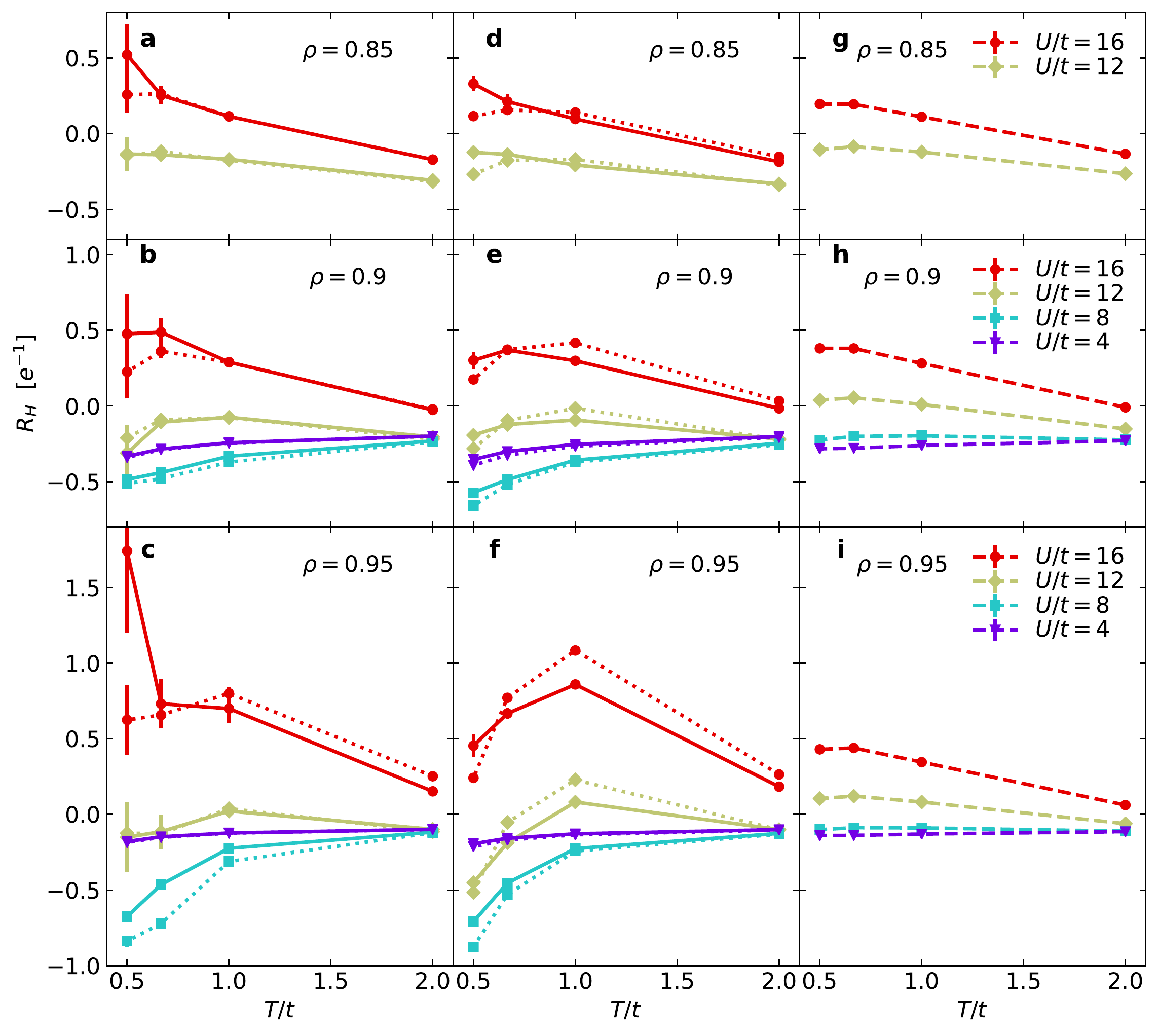}
    \caption{
    dc Hall coefficient as determined from the proxies D, D$_\gamma$, M1, and M2, as well as how these proxies compare to $R_H^{(0)}$.
    M2 has been multiplied by a factor of $4$ to demonstrate that it has similar temperature and doping dependence, as well as sign change structure, as other proxies, but with a reduced magnitude.
    Panels in the same row share the same legend. 
    For all proxies, $\chi_{xy}^{\text{ZF}}(\tau)$ is measured on a $6\times 6$ lattice, and $\chi_{xx}(\tau)$ is measured on a $8\times 8$ lattice.
    Pairs of a dotted and solid lines in panels \textbf{a-f} share the same parameters.  
    \textbf{a-c} Proxy D (solid) and proxy D$_\gamma$ (dotted).
    \textbf{d-f} Proxy M1 (solid) and proxy M2 ($\times 4$, dotted). 
    \textbf{g-i} Dashed lines are $R_H^{(0)}$, defined by the leading order term in the expression for $R_H$ in Refs.~\cite{assa} and \cite{assa2}, constructed from thermodynamic susceptibilities. These are evaluated on $8 \times 8$ lattices. Error bars represent $\pm 1$ standard error determined by jackknife resampling.
}
    \label{fig:RH}
\end{figure*}

We consider two types of proxies that we derive and discuss below:
\begin{itemize}
\item D type -- stemming from an analogy to Drude theory: 
 expressed as
\begin{align}
    R_H^{\mathrm{D}} =  -i T \frac{(\partial_\tau\chi^{\text{ZF}}_{xy})(\tau=\beta/2)}{\left[\chi_{x x}(\tau = \beta/2)\right]^2}, \label{proxy1}
\end{align}
which is obtained by inserting the Drude formulas
\begin{align}
\sigma_{xy}|_{B=0} &=\Omega_{xy}  \frac{1}{(\gamma-\mathit{i}\omega)^2}, \label{xy} \\
\sigma_{xx}|_{B=0}&=\Omega_{xx} \frac{1}{\gamma-\mathit{i}\omega}
    \label{xx}
\end{align}
into Eq.~\ref{Imrelation} and \ref{Rerelation} and taking the limit $\gamma \rightarrow 0$, where $\gamma$ is the scattering rate. 
Another candidate proxy D$_\gamma$ is constructed by assuming $\gamma$ to be non-zero and fitting $\chi_{xx}$ and $\chi_{xy}^{\text{ZF}}$ using Eqs.~\ref{Imrelation},~\ref{Rerelation},~\ref{xy}, and~\ref{xx}. Results of proxies D and D$_\gamma$ are shown in \cref{fig:RH}\textbf{a-c}.
\item M type -- 
determined by extracting the zero Matsubara frequency limit of: 
 \begin{equation}
    R_H^{\mathrm{M1}}(\mathit{i}\omega_n)= \frac{\chi_{xy}^{\text{ZF}}(\mathit{i}\omega_n) \omega_n }{ \l[\chi_{xx}(\mathit{i}\omega_n)-\chi_{xx}(\mathit{i}\omega=0)\r]^2 }, \label{eq:proxym1}
\end{equation}
where $\chi_{\alpha \beta}(\mathit{i}\omega_n)$ in Matsubara frequency is defined as the Fourier transform of the imaginary time data, given by 
$
  \chi_{\alpha \beta}(\mathit{i}\omega_n)  =\int_0^{\beta} \mathrm{d}\tau e^{\mathit{i}\omega_n\tau}\chi_{\alpha \beta}(\tau).
$
Here we utilize a cubic spline interpolant. Another candidate proxy M2 is defined as 
\begin{equation}
    R_H^{\mathrm{M2}}= \frac{\chi_{xy}^{\text{ZF}}(\mathit{i}\omega_1) \pi ^2 }{\omega_1 \l[\chi_{xx}(\tau = \beta/2)\r]^2 \beta^4}, \label{eq:proxym2}
\end{equation}
where $\omega_1 = 2\pi/\beta$, the smallest nonzero Matsubara frequency. 
Results of proxies M1 and M2 are shown in \cref{fig:RH}\textbf{d-f}.
\end{itemize}
Figure~\ref{fig:RH} displays comparisons between $R_H^{(0)}$ and the proxies D, D$_\gamma$, M1, M2, and generally shows that 
D, D$_\gamma$, and M1 closely match $R_H^{(0)}$ from previous work~\cite{wen},
at least within a factor of order unity. Proxy M2, on the other hand, consistently produces results that are much smaller (about $1/4$ of M1) in magnitude.
Calculation details of these proxies are in Appendix~\ref{sec:calcdetail}.

For a Fermi liquid with a momentum-independent scattering rate $\gamma$, theoretical work~\cite{FLomega1,FLomega2,PhysRevLett.84.3418,FLomega4,FLomega5} has demonstrated a Drude-like $\omega$ dependence of the conductivities in Eq.~\ref{xy} and \ref{xx} for $\omega \lesssim \gamma$.
Therefore, we expect proxy D to be in good agreement with the true $R_H$ for a Fermi liquid where the scattering is isotropic and weak. 
Details of the derivation of Eq.~\ref{proxy1} can be found in Appendix~\ref{sec:calcdetail}.

Figures \ref{fig:RH}\textbf{a-c} show proxy D.  
For weak $U$ ($\sim 4t$), or higher doping ($\rho\leq 0.9$), proxy D fits previous results of $R_H^{(0)}$ very well, as expected for a normal metal. 
In these parameter regimes, results for either proxy D or $R_H^{(0)}$ should be deemed reliable.
Approaching the Mott insulator where one should not expect Drude theory to apply, proxy D unsurprisingly deviates from $R_H^{(0)}$.

Proxy D$_\gamma$ assumes $\gamma$ to be  non-zero and finite. 
As shown in Fig.~\ref{fig:RH}\textbf{a-c}, the results for proxy D$_\gamma$ are largely the same as those of proxy D, implying that under the assumptions of Drude theory, our estimation of $R_H$ is not sensitive to changes in relaxation rate. As we can see from Eq.~\ref{xy} and Eq.~\ref{xx}, any direct effect of $\gamma$ cancels out in $R_H$.

There are some limitations for D-type proxies. Conductivities can deviate significantly from Drude theory for strongly interacting systems, leaving the approximate $R_H$ far from the true results. In addition, from solely Eq.~\ref{Imrelation} and \ref{xy}, one can never obtain a $\Im \chi_{xy}(\tau)$ that changes sign as a function of $\tau$ in the range $\tau \in [0,\beta/2)$, which is an important feature in our $\Im \chi_{xy}(\tau)$ data  due to interaction effects.

Now we switch to the M type proxies.
Proxy M1 using Eq.~\ref{eq:proxym1} is exact for dc $R_H$ in the zero-temperature limit $T\ll \Lambda$ ($\omega_n \ll \Lambda$),
where $\Lambda$ is defined as the scale on which $\chi_{\alpha \beta}$ begins to deviate from its low-frequency behavior.
This is because
\begin{align}
    \frac{\sigma_{xy}(i\omega)}{B} &= \frac{\chi_{xy}(\mathit{i}\omega)}{\omega}, \label{imaginaryfrequency} \\
    \sigma_{xx}(i\omega) &= \frac{\chi_{xx}(\mathit{i}\omega)-\chi_{xx}(\mathit{i}\omega=0)}{\omega} \label{imaginaryfrequencyxx}
\end{align}
(see Appendix~\ref{sec:conductivity} and \cite{PhysRevLett.74.3868}).

Proxy M2 approximates $\sigma_{xx}(\omega=0)$ with 
\begin{equation}
-\frac{\beta^2}{\pi} \chi_{xx}(\tau = \beta/2) \label{proxy_sigxx}
\end{equation}
and uses $\omega = \omega_1$ to estimate Eq.~\ref{imaginaryfrequency} at $\omega=0$.
Equation \ref{proxy_sigxx} is also exact at $T\ll \Lambda$,
and is often used as a proxy for $\sigma_{xx}$ in other work~\cite{huang}. 

\begin{figure}[htbp]
    \centering
    \includegraphics[width=0.95\textwidth]{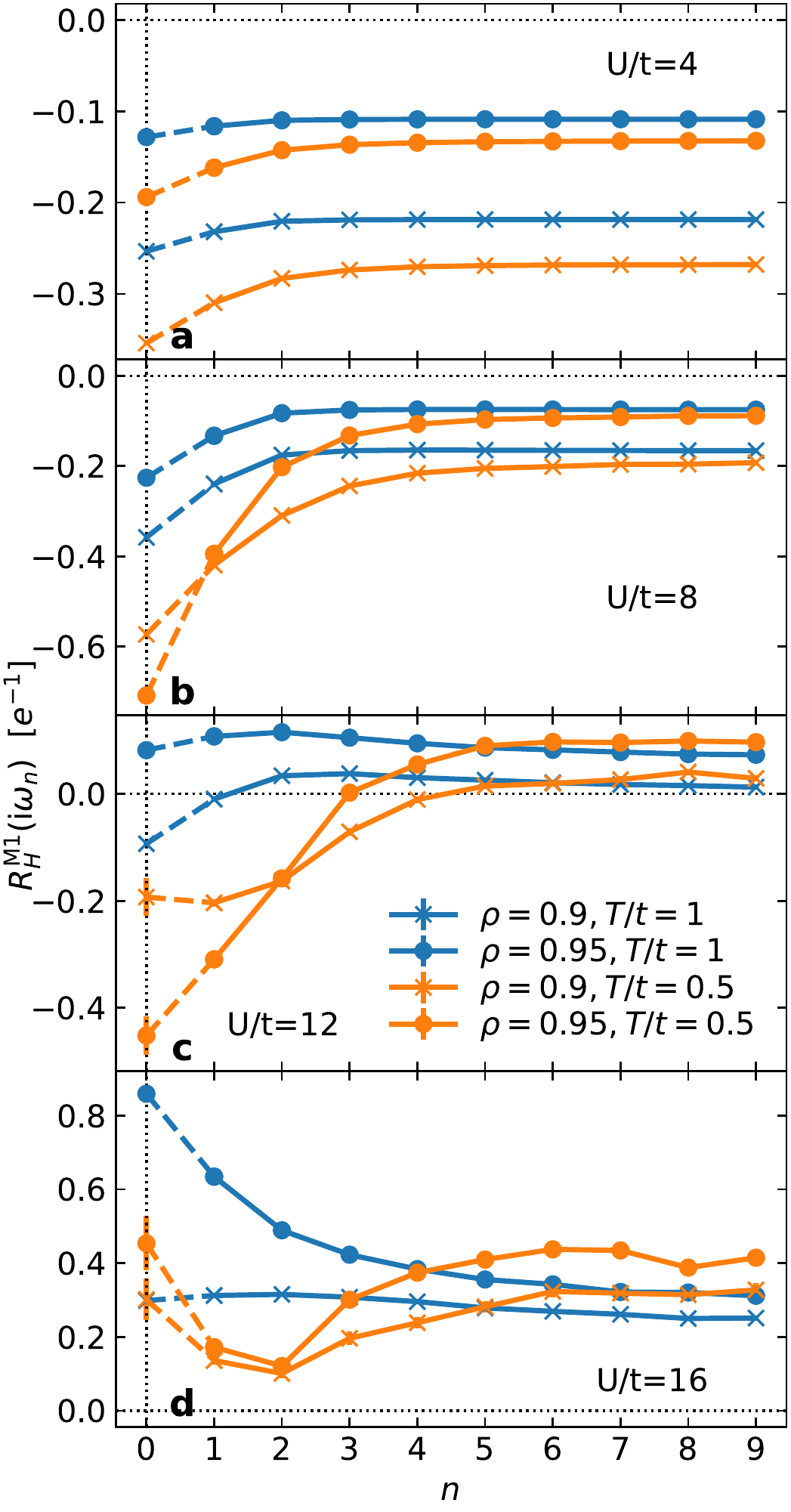} 
    \caption{$R_H^{\mathrm{M1}}(\mathit{i}\omega_n)$ in Matsubara frequency. 
    Data are shown for $U/t=4 - 16$ for panels \textbf{a-d}, respectively, with $\rho=0.9$ and $0.95$ and temperature $T/t = 0.5$ and $1$. 
    The data points at $\omega_n=0$ correspond to proxy M1 shown in Fig.~\ref{fig:RH}. 
    $\chi_{xy}^{\text{ZF}}$ was calculated on a $6\times 6$ lattice, while
    $\chi_{xx}$ was calculated on an $8\times 8$ lattice.  Error bars for $n>0$ are smaller than the points.
    }
    \label{matsuraw}
\end{figure}

Figures \ref{fig:RH}\textbf{d-f} show proxies M1 and M2. Values of M1 are overall close to those for both proxy D and D$_\gamma$, even for strong interactions up to $U=16\,t$.
Indeed generally, as long as the rapidly varying components in $\sigma_{xy}(i\omega)$ and $\sigma_{xx}(i\omega)$ cancel out in $R_H^{\mathrm{M1}}(i\omega)$, proxy M1 is very accurate regardless of the explicit form of frequency dependence of the conductivities. 
In other words, proxy M1 only requires the ratio of the conductivities, $R_H^{\mathrm{M1}}(i\omega_n)$, to vary slowly with $\omega_n$, as we show in Fig.~\ref{matsuraw}, where we plot $R_H^{\mathrm{M1}}(\mathit{i}\omega_n)$ against $n$ (similar to Ref.~\cite{PhysRevLett.74.3868}). Proxy M2 is not able to make use of cancellation between the transverse and longitudinal conductivities, so it still requires $T\ll \Lambda$ for both conductivity components.
The difference between values of M1 and M2 in our simulations indicates that $\Lambda \lesssim T$, so that $\sigma_{xy}(i\omega)$ and $\sigma_{xx}(i\omega)$ vary significantly within the scale set by the smallest non-zero Matsubara frequency $\omega_1$.

Considering that D and M type proxies have different assumptions and approach $R_H$ from quite different aspects, it is remarkable that D, D$_\gamma$, and M1 are overall comparable to each other. 
Even more remarkably, all the proxies 
compare particularly well to the sign changing structure to $R_H^{(0)}$. 
Also considering the previous comparison of ZF results with those of FF in Sec.~\ref{3currderivations}, 
we conclude that the previous method we used to calculate $R_H^{(0)}$~\cite{wen} as an approximation for $R_H$ is reliable for large $U$, despite our neglect of higher order corrections. 
As we have shown in this section, the proxies constructed using $\chi_{xy}^{\text{ZF}}$ and $\chi_{xx}$ are also a useful approximation when direct analytic continuation to find $\sigma_{xy}(\omega)$ is challenging.

We remark that within the temperatures accessible for DQMC simulation, our $R_H$ results do not show a strong doping dependence away from half-filling, specifically in the region around $20 \%$ hole doping as referenced in cuprate experiments~\cite{badoux2016change,PhysRevB.95.224517,doiron2017pseudogap}. 

\section{discussion}\label{discussion}

Through a consideration of various methods to evaluate the Hall coefficient, we have demonstrated agreement
between the sign change structures of $\chi_{xy}^{\text{ZF}}$, $\chi_{xy}^{\text{FF}}$, and $R_H^{(0)}$, supporting prior claims that the sign of the dc Hall coefficient has a close relationship to the Fermi surface topology in the strongly correlated zero-field Hubbard model~\cite{wen}. 
The rough agreement between proxy D, D$_\gamma$, M1 and $R_H^{(0)}$, as well as the significantly different result of proxy M2, suggests that making use of cancellation between transport quantities may simplify evaluations of difficult multifermion correlation functions such as the Hall coefficient and allow us to construct a good description of transport without quasiparticles.
The following facts lend support to our idea.
While longitudinal resistivity in the Hubbard model shows typical non-Fermi liquid behavior~\cite{huang}, $R_H$ shows relatively flat $\omega_n$-dependence in Fig.~\ref{matsuraw}.
 In this work, proxy D and D$_\gamma$ provide similar results. This likely results from our assumption that $\gamma$ is the same for the two conductivities, leading indirectly to simplifications that allow D$_\gamma$ to mimic D due to apparent cancellation of lifetime effects.
 In constructing proxy M2, there are no such cancellations, and 
the proxy is fragile and easily fails.
Finally, Refs.~\cite{assa} and~\cite{assa2} showed how ratios of conductivities like the Hall coefficient or thermal Hall coefficient reduce to expressions constructed from simple thermodynamic susceptibilities.
It is an open and intriguing question whether
a Fermi-liquid like correspondence between Fermi surface topology and $R_H$ in Ref.~\cite{wen} is due to such cancellations between conductivities, rather than necessarily Fermi-liquid-like $\omega$-dependence of each conductivity.

It remains a challenge to perform analytic continuation directly from Matsubara frequencies or imaginary time data. 
A promising approach would be to use techniques designed to treat non-positive-definite spectra~\cite{PhysRevB.92.060509,PhysRevB.95.121104}, or other methods of analytic continuation~\cite{fei2020nevanlinna,PhysRevB.98.205102,PhysRevLett.111.182003, PhysRevD.95.056016}.
If a reliable method of analytic continuation can be found, then our evaluation of the exact three-current linear-response $\chi_{xy}^{\text{ZF}}$ through the numerically exact and unbiased DQMC algorithm will allow us to find the exact $\sigma_{xy}(\omega)$ spectra for all frequencies for the Hubbard model. 
Even in the absence of reliable analytic continuation methods, our $\chi_{xy}^{\text{ZF}}$ results will still be a benchmark for any theory that proposes a frequency dependence of $\sigma_{xy}$ for the strongly correlated Hubbard model or similar models. 

The data and analysis routines (Jupyter/Python) needed to reproduce the figures can be found at  \url{https://doi.org/10.5281/zenodo.4569163}.

\begin{acknowledgments}
We acknowledge helpful discussions with S. Kivelson, S. Lederer, A. Auerbach and I. Khait.
\emph{Funding:}
This work was supported by the U.S. Department of Energy (DOE), Office of Basic Energy Sciences,
Division of Materials Sciences and Engineering. 
EWH was supported by the Gordon and Betty Moore Foundation EPiQS Initiative through the grants GBMF 4305 and GBMF 8691.
YS was supported by the  Gordon and Betty Moore Foundation’s EPiQS Initiative through grants  GBMF 4302  and  GBMF 8686.
Computational work was performed on the Sherlock cluster at Stanford University and on resources of the National Energy Research Scientific Computing Center, supported by the U.S. DOE, Office of Science, under Contract no. DE-AC02-05CH11231.
\end{acknowledgments}

\appendix

\section{simulation details}\label{sec:simulationdetails}

We use DQMC simulations to measure  $\chi_{xy}^{\text{ZF}}(\tau)$ for the Hubbard model in Eq.~\ref{eq:hubbard} without $B$ and $\chi_{xy}^{\text{FF}}(\tau)$ under $eB=2\pi/V$,  respectively, on a 2D square lattice with the corresponding periodic boundary conditions~\cite{PhysRevB.65.115104}. Strategies for evaluating Green's functions in this work are the same as those of Ref.~\cite{huang}.

We can write down an alternate expression for $\chi_{xy}^{\text{ZF}}$:
\begin{align}
\chi_{xy}^{\text{ZF}}(\tau)
    =&\int_0^\beta \dd\tau'\sum_{\mathbf{r}'\mathbf{r}''}\frac{\sin(q(x'-x''))}{q} \nonumber \\&\qquad\times\langle T_{\tau} J_{y}(\mathbf{r}'',\tau)J_{y}(\mathbf{r}',\tau')J_{x}(\mathbf{0},0)\rangle. \label{eq:3currentanother}
\end{align} 
Because of $C_4$ symmetry, we confirm that $\chi_{xy}^{\text{ZF}}$ given in Eqs.~\ref{eq:3current} and \ref{eq:3currentanother} give identical results and report the average of the two quantities to reduce sampling errors and improve statistical measurements.
Similarly, since $\chi_{xy}^{\text{FF}}(\tau) =  -\chi_{yx}^{\text{FF}}(\tau)$, we measure and report $ (\chi_{xy}^{\text{FF}} - \chi_{yx}^{\text{FF}} )/ 2$. 

In our simulations for $\chi_{xy}^{\text{ZF}}$, we measure $\expval{T_{\tau}\mathbf{J}(\tau)\mathbf{J}(\tau ')\mathbf{J}(0)}$
at discretized $\tau '$ values.
The integral over $\tau '$  is  computed by  finding two cubic spline interpolants using data points in the ranges $\tau '\in[0,\tau]$ and $\tau '\in[\tau,\beta]$ respectively (here $\expval{T_{\tau}\mathbf{J}(\tau)\mathbf{J}(\tau ')\mathbf{J}(0)}$ is discontinuous at $\tau'=\tau$ due to time ordering) and then integrating the interpolants.

\begin{figure}
    \centering
    \includegraphics[width=0.9\textwidth]{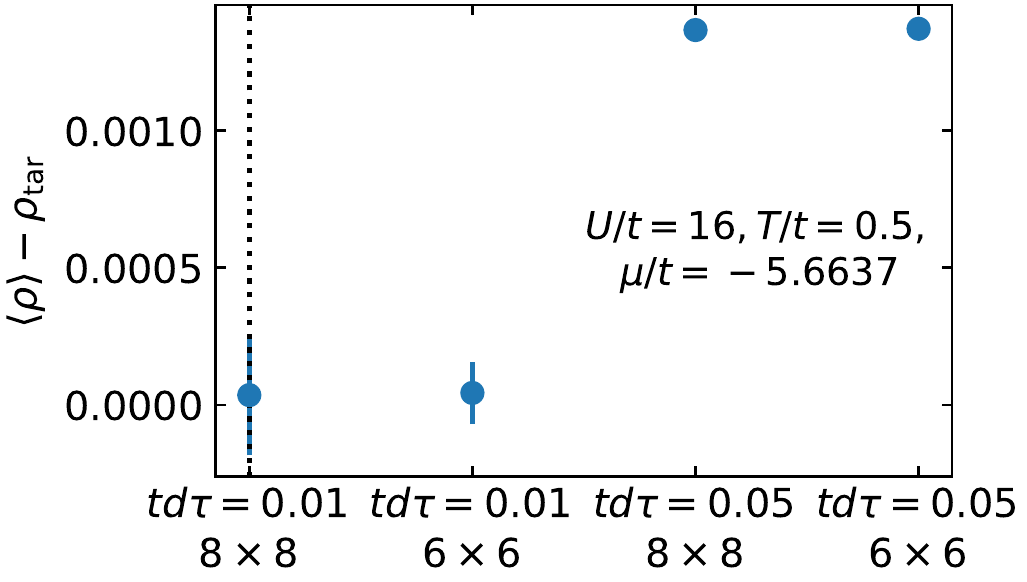}
    \caption{An example of the error in density associated with $\mu$ for the ZF case, shown as the difference between the measured density $\ev{\rho}$ and the target density $\rho_{\text{tar}}$, for different $\dd\tau$ and lattice sizes. The parameters $U/t=16$, temperature $T/t=0.5$ and $\rho_{\text{tar}}=0.9$ are selected, as $U/t=16$ is the largest interaction strength that we consider, with the most severe Trotter error. The dotted line indicates the parameters used in tuning $\mu$ for the target $\rho_{\text{tar}}$ (The largest cluster and smallest $\dd\tau$). 
    }
    \label{densityerror}
\end{figure}

To tune the chemical potential $\mu$ for a specific target filling level $\rho_{\text{tar}}$ at a specific temperature, we use DQMC to calculate $\ev{\rho}$ for a range of chemical potentials $\mu$ (at $0.05t$ intervals for ZF case and $0.1t$ intervals for FF case) and obtain the best $\mu$ by linear interpolation of the $\ev{\rho}$ versus chemical potential curve. In the ZF case, $\mu$ is tuned on a $8\times 8$ lattice  and the same $\mu$ is used for both  $8 \times 8$ and $6 \times 6$ lattice sizes. In the FF case, $\mu$ tuning for $6 \times 6$ and $8 \times 8$ lattices are done separately, although in practice, the optimal $\mu$ is almost identical for the two lattice sizes.
For our parameters, we can obtain $\ev{\rho}$ to within a tolerance of $O(10^{-3})$ of the target density $\rho_{\text{tar}}$ (written as $\rho$ throughout this work).  
An example of this tolerance in the ZF case is shown in Fig.~\ref{densityerror}.
Including the effects of the Trotter error, DQMC statistical error, and the density shift between lattice sizes for specific values of $\mu$, our density is accurate to $O(10^{-3})$.

\begin{figure*}
    \centering
    \includegraphics[width=0.9\textwidth]{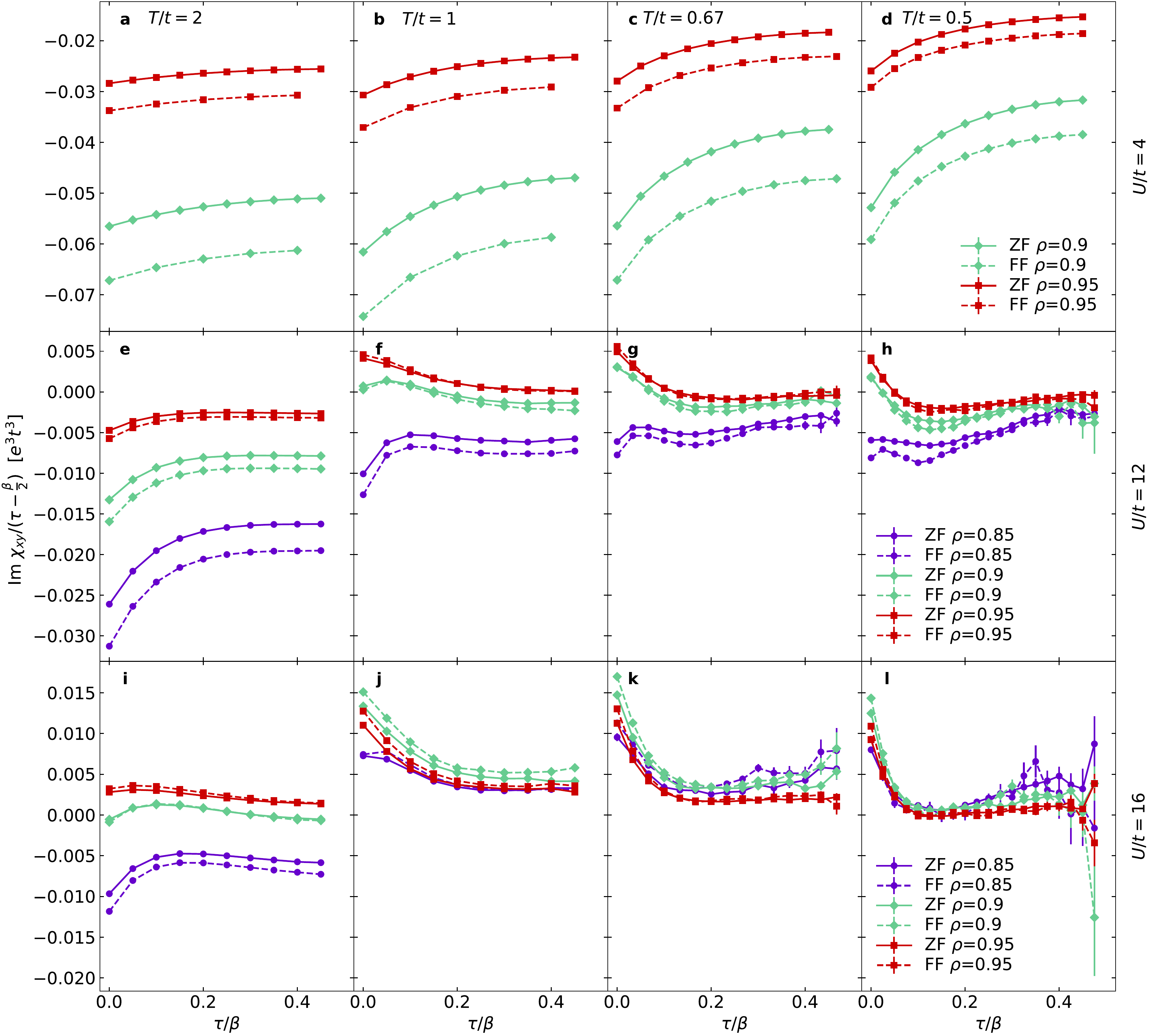}
    \caption{Supplementary data for Fig.~\ref{fig:compare_finitefield}
    with a direct comparison between the ZF $\Im \chi_{xy}^{\text{ZF}}(\tau)$ and FF $\Im \chi_{xy}^{\text{FF}}(\tau)$.  
    Calculations were performed on $6\times6$ lattices using the DQMC algorithm. 
    Error bars are $\pm$ 1 standard error determined by  jackknife resampling. 
    The same color (marker) within each panel represents the same parameters, as referenced in the legends in the final column. Rows have fixed interaction strength $U$ and columns have the fixed temperature $T$.
    }
    \label{fig:morefffzcompare}
\end{figure*}

Regarding the Trotter error, we define $\dd\tau$ as the interval between imaginary time data points. 
In previous work~\cite{wen}, we set a minimum partition of imaginary time $L=\beta/\dd\tau= 20$ and a maximum $\dd\tau = 0.1/t$ 
for all interactions and temperatures.
In this work, for $U/t=4 - 8$, all ZF calculations ($\chi_{xy}^{\text{ZF}}$, $\chi_{xx}$, $R_H^{(0)}$, and  $\ev{\rho}$) use the same $\dd\tau$ as in the previous work~\cite{wen}, while FF calculations ($\chi_{xy}^{\text{FF}}$ and  $\ev{\rho}$) 
also have maximum $\dd\tau = 0.1/t$, but have minimum $L=10$. So the imaginary time spacing of $\chi_{xy}^{\text{FF}}$ is larger than $\chi_{xy}^{\text{ZF}}$ at the highest temperatures, as shown in Fig.~\ref{fig:morefffzcompare}\textbf{a}-\textbf{d} for $U/t=4$ and Fig.~\ref{fig:finitesizeU8} for $U/t=8$. 
For $U/t=12 - 16$, $\chi_{xy}^{\text{ZF}}$, $\chi_{xx}$, $\chi_{xy}^{\text{FF}}$, and $R_H^{(0)}$ are all obtained with a maximum $\dd\tau = 0.05/t$ and minimum $L= 20$, in order to reduce the Trotter error.  Measurement of $\ev{\rho}$ in tuning of $\mu$ is done using a minimum $L= 20$, and a maximum $\dd\tau=0.01/t$ (ZF) and $\dd\tau=0.02/t$ (FF).
In summary, $Ut(\dd\tau)^2 \leq 0.08$ in this work, comfortably below the conventionally adopted limit $Ut(\dd\tau)^2 \leq 1/8$.  We consider Trotter error to be negligible.
The accuracy of our results is affected primarily by the limitations of individual proxies, as discussed in the main text. In addition to Trotter error discussed here, our only other source of systematic error is finite-size effects discussed in Appendix~\ref{sec:finitesize}.

We run up to approximately $600$ independently seeded Markov chains for $\chi_{xy}^{\text{ZF}}$,
about $40$ Markov chains for $\chi_{xx}$ and $R_H^{(0)}$, and up to $200$ Markov chains for $\chi_{xy}^{\text{FF}}$. 
For $\chi_{xy}^{\text{ZF}}$, when $U/t=12-16$, 
each Markov chain has $5\times 10^{5}$ space-time sweeps, while for $U/t=4-8$, 
each Markov chain has $10^{6}$ space-time sweeps.
$\chi_{xx}$ and $R_H^{(0)}$ are measured together and have $10^{6}$ space-time sweeps in each Markov chain.
$\chi_{xy}^{\text{FF}}$ has $5\times 10^{5}$ to $4\times 10^{6}$ space-time sweeps in each Markov chain, depending on parameters.
For $\chi_{xy}^{\text{ZF}}$, $\chi_{xx}$ and $R_H^{(0)}$, measurements are performed once every $4$ sweeps.
For $\chi_{xy}^{\text{FF}}$, measurements are performed once every $2$ sweeps.
Measurements occupy more than $90 \%$ of the simulation runtime, meaning $\chi_{xy}^{\text{ZF}}$ is $O(L_x L_y L)$ more expensive than $\chi_{xy}^{\text{FF}}$.

\section{dynamical conductivity}
\label{sec:conductivity}
In this section we discuss the relationship between $\sigma_{xy}$ and $
\chi_{xy}$.
In the derivations of this section and Appendix~\ref{sec:calcdetail}, that relate  current-current correlation functions to conductivities, we assume the thermodynamic limit. In that limit, $\chi_{xy}^{\text{ZF}}$ is considered to be the linear response to an infinitesimal uniform magnetic field $B$,
\textit{i.e.} $\chi_{xy}^{\text{ZF}} =\lim_{B\rightarrow 0}\chi_{x y}$. On the other hand, in defining $\chi_{xy}^{\text{ZF}}$ in Eq.~\ref{eq:3current}, we used a finite $q$ and a non-uniform magnetic field to ensure that the expression is well-defined on a finite lattice. 
This approach is reasonable, as the conductivities obtained in this way converge to their values in the thermodynamic limit as $q\rightarrow 0$.

With the definition of the total current operator $J_\alpha \equiv \sum_{\mathbf{r}} J_\alpha(\mathbf{r})$, the definitions of the $J_x-J_y$ correlators, in imaginary time $\tau$ and Matsubara frequency $\mathit{i}\omega_n=\mathit{i} 2 \pi n/\beta$, are
\begin{align}
&\chi_{x y}(\tau) = -\frac{1}{V}\ev{J_x(\tau) J_y}_B/B  \nonumber\\
&= -\frac{1}{ZVB}\sum_{p m} \mel{p}{J_x}{m} \mel{m}{J_y}{p} e^{-\beta E_p} e^{\tau (E_p - E_m)}, \label{chitau}\\
&\chi_{x y}(\mathit{i}\omega_n) = \int_0^\beta \dd{\tau} e^{\mathit{i} \omega_n \tau} \chi_{x y}(\tau)   \nonumber \\
&= \frac{1}{ZVB}\sum_{p m} \mel{p}{J_x}{m} \mel{m}{J_y}{p} \frac{e^{-\beta E_p} -  e^{-\beta E_m}}{\mathit{i}\omega_n + E_p - E_m}. \label{chiiomega}
\end{align}
In real-time and real-frequency, 
\begin{align}
&\chi_{x y}^R(t) = -\frac{\mathit{i}}{VB} \theta(t) \ev{\comm{J_x(t)}{J_y}}_B 
= \frac{-\mathit{i}\theta(t)}{ZVB} \sum_{p m}
\nonumber \\ &
 \mel{p}{J_x}{m} \mel{m}{J_y}{p} 
(e^{-\beta E_p} -  e^{-\beta E_m}) e^{\mathit{i} t (E_p - E_m)}, \\
&\chi_{x y}^R(\omega) = \int \dd{t} e^{\mathit{i} \omega t} \chi_{x y}^R(t) \nonumber\\
&= \frac{1}{ZVB}\sum_{p m} \mel{p}{J_x}{m} \mel{m}{J_y}{p} \frac{e^{-\beta E_p} -  e^{-\beta E_m}}{\omega  + E_p - E_m + \mathit{i}0^+}. \label{chiR}
\end{align}
Using $\dfrac{1}{x + \mathit{i}0^+} = \mathcal{P}\dfrac{1}{x} - \mathit{i} \pi \delta(x)$, we can break up Eq.~\ref{chiR},
\begin{align}
\chi_{x y}^R(\omega) &= \chi_{x y}^{(1)}(\omega) + \mathit{i} \chi_{x y}^{(2)}(\omega),  \\
\chi_{x y}^{(1)}(\omega) &= \frac{1}{ZVB}\sum_{p m} \mel{p}{J_x}{m} \mel{m}{J_y}{p} \nonumber \\ &(e^{-\beta E_p} -  e^{-\beta E_m})\mathcal{P}\frac{1}{\omega + E_p - E_m}, \\
\chi_{x y}^{(2)}(\omega) &= \frac{-\pi}{ZVB}\sum_{p m} \mel{p}{J_x}{m} \mel{m}{J_y}{p} \nonumber \\ & (e^{-\beta E_p} -  e^{-\beta E_m})\delta(\omega + E_p - E_m). \label{chi2}
\end{align}
Comparing Eq.~\ref{chitau} and Eq.~\ref{chi2}, we see
\begin{equation}
\chi_{x y}(\tau) = \int \frac{\dd{\omega}}{\pi} \frac{e^{-\tau \omega}}{1 - e^{-\beta \omega}} \chi_{x y}^{(2)}(\omega). \label{eqAC}
\end{equation}

$\chi_{xy}(\tau)$ is purely imaginary, which one sees as follows: 
$\ev{J_x(\tau)J_y}=\ev{J_{y}(\tau)J_{-x}}=-\ev{J_{y}(\tau)J_{x}}$. 
In addition,
\begin{align}
    &\ev{J_y(\tau) J_x}_B =\frac{1}{Z}\mathrm{Tr}( \mathrm{exp}((\tau-\beta) H) J_y \mathrm{exp}(-\tau H) J_x) \nonumber\\ &=\frac{1}{Z}\mathrm{Tr}(J_y \mathrm{exp}(-\tau H) J_x \mathrm{exp}((\tau-\beta) H)) = \ev{J_x(\tau) J_y}_B^*, \nonumber
\end{align}
which gives
\begin{equation}
    \ev{J_x(\tau) J_y}_B^* = -\ev{J_x(\tau) J_y}_B.
\end{equation}
Since $\chi_{x y}^{(2)}(\omega)$ and $\chi_{x y}^{(1)}(\omega)$ are related by a Kramers-Kronig transform, $\chi_{x y}^{(1)}(\omega)$ is also purely imaginary. So the real part of $\chi_{xy}^R(\omega)$ is $\mathit{i}\chi_{x y}^{(2)}(\omega)$ and the imaginary part of $\chi_{xy}^R(\omega)$ is $\chi_{x y}^{(1)}(\omega)$.
The Hall conductivity is (by Kubo formula~\cite{mahan2013many})
\begin{align}
\frac{\sigma_{x y}(\omega)}{B} &= \frac{\chi_{x  y}^R(\omega)}{-\mathit{i} \omega}, \label{sigmaxyoriginal} \\
\frac{\Re \sigma_{x y}(\omega)}{B} &=- \frac{\chi_{x y}^{(1)}(\omega)}{\mathit{i} \omega}, \label{resigxy}
\\
\frac{\Im \sigma_{x y}(\omega)}{B} &= \frac{\mathit{i}\chi_{x  y}^{(2)}(\omega)}{ \omega}. \label{imsigxy}
\end{align}
Combining Eq.~\ref{imsigxy} with Eq.~\ref{eqAC}, we obtain the useful relation
\begin{align}
    \frac{\chi_{x y}(\tau)}{\mathit{i}} = -\int \frac{\dd{\omega}}{\pi} \frac{e^{-\tau \omega}}{1 - e^{-\beta \omega}} \omega \frac{\Im \sigma_{xy}(\omega)}{B}. \label{Imrelation0}
\end{align}
The analogous relation we have employed previously to study the diagonal conductivity by measuring $\chi_{xx}\equiv -\ev{J_x(\tau)J_x}_B/V$ is~\cite{huang}
\begin{align}
    \chi_{x x}(\tau) = -\int \frac{\dd{\omega}}{\pi} \frac{e^{-\tau \omega}}{1 - e^{-\beta \omega}} \omega \Re \sigma_{xx}(\omega). \label{Rerelation0}
\end{align}

$\Im \sigma_{xy}(\omega)/\omega$ and $\Re \sigma_{xx}(\omega)$ are even functions, so we can transform Eq.~\ref{Imrelation0} and Eq.~\ref{Rerelation0} such that
\begin{align}
    \frac{\chi_{x y}(\tau)}{\mathit{i}} &= -\int_0^{\infty} \frac{\dd{\omega}}{\pi} \frac{\omega^2 \left[e^{(\frac{\beta}{2}-\tau) \omega}-e^{-(\frac{\beta}{2}-\tau) \omega}\right]}{
    e^{\frac{\beta}{2} \omega} - e^{-\frac{\beta}{2} \omega}} \frac{\Im \sigma_{xy}(\omega)}{\omega B},  \nonumber\\
    \chi_{x x}(\tau) &= -\int_0^{\infty} \frac{\dd{\omega}}{\pi} \frac{\omega \left[e^{(\frac{\beta}{2}-\tau) \omega}+
    e^{-(\frac{\beta}{2}-\tau) \omega}\right]}{e^{\frac{\beta}{2} \omega} - e^{-\frac{\beta}{2}\omega}}  \Re \sigma_{xx}(\omega).\nonumber 
\end{align}
These are Eq.~\ref{Imrelation} and \ref{Rerelation} in the main text.

Comparing Eq.~\ref{chiiomega} and Eq.~\ref{chiR} and considering Eq.~\ref{sigmaxyoriginal},
we also conclude that
\begin{align}
    \frac{\sigma_{xy}(\omega = 0)}{B} &=\lim_{\omega\rightarrow 0 } \frac{\chi_{xy}(\mathit{i}\omega)}{\omega}. 
\end{align}

\section{Finite size effects}\label{sec:finitesize}
\begin{figure}
    \centering
    \includegraphics[width=\textwidth]{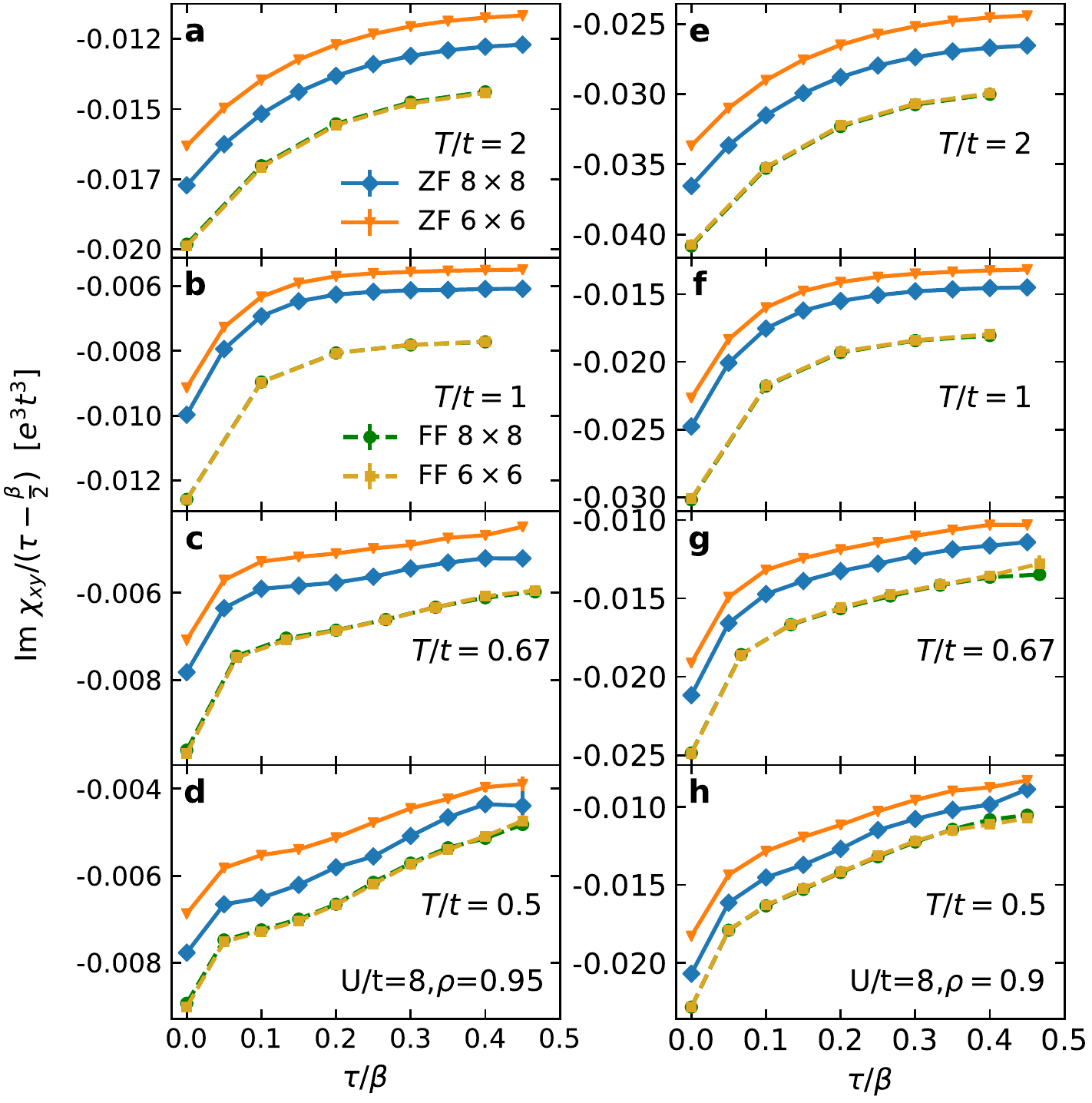}
    \caption{Finite size analysis for $\Im \chi_{xy}^{\text{ZF}}$ and $\Im \chi_{xy}^{\text{FF}}$ in imaginary time with $U/t=8$ and $T/t=0.5 - 2$.  $\Im\chi_{xy}^{\text{ZF}}(\tau)$ (solid) and $\Im\chi_{xy}^{\text{FF}}(\tau)$ (dashed) are divided by $\tau-\frac{\beta}{2}$ to accentuate the behavior near $\tau=\beta/2$.
    \textbf{a-d} Filling $\rho=0.95$ and \textbf{e-h} filling $\rho=0.9$.
    }
    \label{fig:finitesizeU8}
\end{figure}

\begin{figure}
    \centering
    \includegraphics[width=0.65\textwidth]{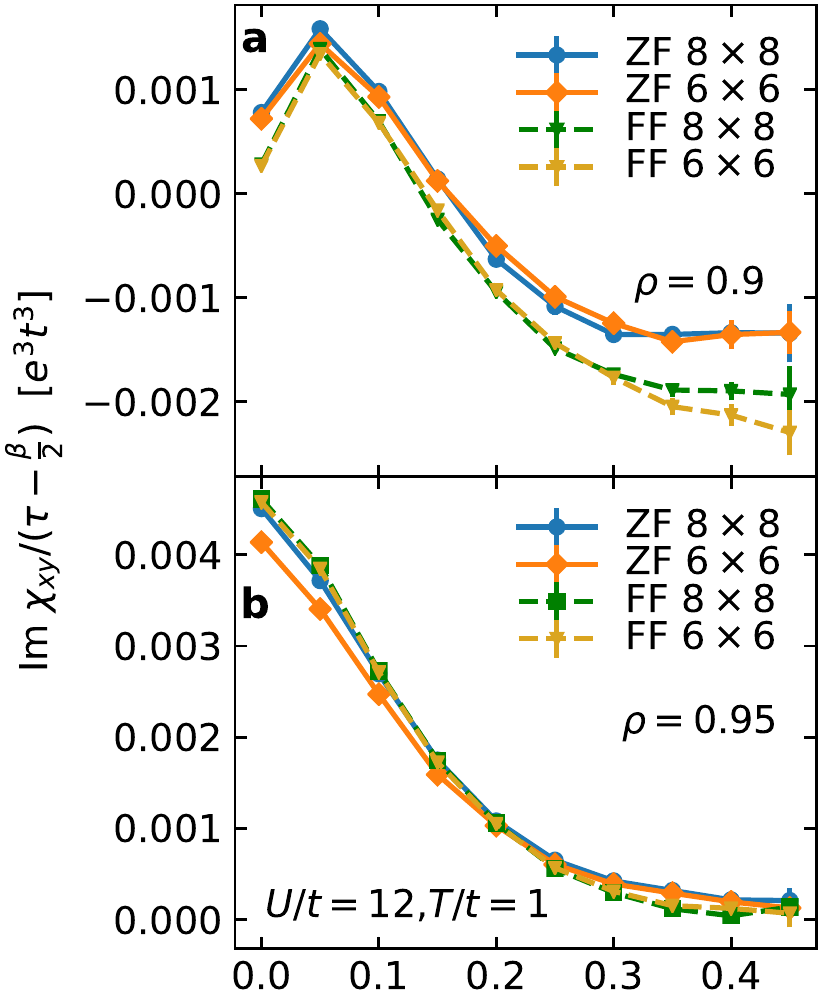}
    \caption{Finite size analysis for $\Im \chi_{xy}^{\text{ZF}}$ and $\Im \chi_{xy}^{\text{FF}}$ in imaginary time for $U/t=12$ and $T/t=1$.}
    \label{fig:finitesizeU12}
\end{figure}

A comparison of ZF and FF data obtained using different lattice sizes for $U/t=8$ and $U/t=12$ is shown in Fig.~\ref{fig:finitesizeU8} and \ref{fig:finitesizeU12}, respectively.
Fig.~\ref{fig:finitesizeU8} reveals that the FF results have smaller finite-size effects than the ZF results.  As we have discussed in the main text, we expect the ZF and FF  results to be identical in the thermodynamic limit, and that trend is indeed shown in Fig.~\ref{fig:finitesizeU8} and Fig.~\ref{fig:finitesizeU12}.

\section{proxy calculation details}\label{sec:calcdetail}
In this section, we start from the Drude formula in Eq.~\ref{xy} and Eq.~\ref{xx} 
and derive proxy D in the limit of weak scattering $\gamma \rightarrow 0$.
From Eq.~\ref{xy} and using $\lim_{\gamma \rightarrow 0}\dfrac{2\omega^2\gamma}{\pi(\gamma^{2}+\omega^2)^2}=\delta(\omega)$,
we have 
\begin{align}
    \lim_{\gamma \rightarrow 0} \omega\Im \sigma_{xy} 
    &=\Omega_{xy} \pi \delta(\omega). 
\end{align}
Inserting this into Eq.~\ref{Imrelation} and calculating the derivative,
\begin{align}
    &\lim_{\gamma \rightarrow 0} \left.\frac{\partial }{\partial \tau}\frac{\chi_{x y}(\tau)}{\mathit{i}} \right|_{\tau = \beta/2} \nonumber \\
    &=  \int_{-\infty}^{\infty} \frac{\dd{\omega}}{\pi} \frac{\omega}{e^{\beta \omega/2} - e^{-\beta \omega/2}} \omega \lim_{\gamma \rightarrow 0} \Im \sigma_{xy}(\omega)/B \nonumber \\
    &= \frac{1}{\beta B}\Omega_{xy}. \label{deri}
\end{align}
From Eq.~\ref{xx} and considering $\lim_{\gamma \rightarrow 0}\frac{\gamma}{(\gamma^{2}+\omega^2)}=\pi \delta(\omega)$,
\begin{align}
    \lim_{\gamma \rightarrow 0} \Re \sigma_{xx} 
    &=\Omega_{xx} \pi \delta(\omega). 
\end{align}
Inserting this into Eq.~\ref{Rerelation} we obtain
\begin{align}
    &\lim_{\gamma \rightarrow 0} \chi_{x x}(\tau = \beta/2)  \nonumber\\&=  -\int_{-\infty}^{\infty} \frac{\dd{\omega}}{\pi} \frac{1}{e^{\beta \omega/2} - e^{-\beta \omega/2}} \omega \lim_{\gamma \rightarrow 0} \Re \sigma_{xx}(\omega) \nonumber \\
    &= -\frac{1}{\beta}\Omega_{xx}. \label{noderi}
\end{align}
So combining Eqs.~\ref{xy},~\ref{xx},~\ref{deri}, and \ref{noderi}, we finally have the expression for the dc Hall coefficient under the assumption that $\gamma\rightarrow 0$,
\begin{align}
    R_H =  \l.\left[\frac{\partial }{\partial \tau}\frac{\chi_{x y}(\tau)}{\mathit{i}}\right] \r|_{\tau = \beta/2} / (\beta  \l[\chi_{x x}(\tau = \beta/2)\r]^2 ).
\end{align}

One can test that for the Hubbard model  with $U=0$, $\chi_{xx}(\tau)$ does not vary with $\tau$, and  $\partial_\tau \chi_{xy}^{\text{ZF}}(\tau)$ does not change with $\tau$ in the thermodynamic limit ($a/L_x$ and $a/L_y \ll T/t$).
This implies that $\omega\Im \sigma_{xy}(\omega)$ and $\Re \sigma_{xx}(\omega)$ are both $\propto \delta(\omega)$.

For proxy D, we use finite differences to estimate $(\partial_\tau \chi_{xy}^{\text{ZF}})(\tau=\beta/2)$. Error bars for proxy D are constructed by error propagation of the standard errors in $(\partial_\tau \chi_{xy}^{\text{ZF}})(\tau=\beta/2)$ and $\left[\chi_{x x}(\tau = \beta/2)\right]^2$, which themselves are determined by jackknife resampling. 

For proxy D$_\gamma$, 
We fit a few values of  $\chi_{xy}^{\text{ZF}}$ and $\chi_{xx}$  near $\tau = \beta/2$ to Eq.~\ref{xy} and ~\ref{xx} through Eq.~\ref{Imrelation} and \ref{Rerelation}.
In doing so, we assume that the $\gamma$s are equal in $\sigma_{xx}$ and $\sigma_{xy}$. We also make use of the fact that values of $\chi_{xx}(\tau)$ and $\chi_{xy}(\tau)$ near $\tau=\beta/2$ are determined more predominately by the low-frequency Drude-like behavior of conductivities, compared with other $\tau$ values.
By fitting $\chi_{xx}$ we extract $\gamma$ and $\Omega_{xx}$; and by fitting $\chi_{xy}^{\text{ZF}}$ we extract $\Omega_{xy}/B$.

We choose to use $\sigma_{xx}$ rather than $\sigma_{xy}$ to find $\gamma$ for several reasons:  DQMC measurements of $\chi_{xx}(\tau)$ have smaller numerical errors than $\chi_{xy}^{\text{ZF}}(\tau)$, and we could use the second derivative of $\chi_{xx}$ to estimate $\gamma$, but we need to use the third derivative of $\chi_{xy}$ to estimate $\gamma$. Correspondingly, using $\chi_{xy}^{\text{ZF}}$ would have required us to fit more data points around $\beta/2$ to calculate $\gamma$.
In general, we need to use at least two data points for the $\chi_{xx}$ fit, and at least one data point for the $\chi_{xy}^{\text{ZF}}$ fit. When numerical errors are large, we choose to include more data points in order to obtain a accurate fit. The downside of including points further away from $\tau=\beta/2$ is that we must assume Drude-type behavior holds on a wider frequency range for $\sigma_{xx}$ and $\sigma_{xy}$.
Note that $\chi_{xx}(\tau)$ and $\chi_{xy}^{\text{ZF}}(\tau)$ are symmetric and antisymmetric about $\tau = \beta/2$, respectively, so we only use data points on one side of $\tau = \beta/2$.

Regarding the proxy D$_\gamma$,
 for $U/t \leq 8$, only the points at $\tau = \beta/2$ and $\tau = \beta/2-\beta/L$ are used in the fitting procedure, while For $U/t \geq 12$, points at $\tau = \beta/2$,$\tau = \beta/2-\beta/L$ and $\tau = \beta/2-2\times \beta/L$ are used. 
    The error in $\gamma$ and $\Omega_{xx}$ obtained from fitting $\chi_{xx}$ is neglected, because $\chi_{xx}$ has much smaller relative error than $\chi_{xy}^{\text{ZF}}$.
    The error in $\Omega_{xy}/B$ obtained from fitting $\chi_{xy}^{\text{ZF}}$ is $\pm 1$ standard error determined by jackknife resampling.
    
For the M-type proxy, when we calculate $\chi_{\alpha \beta}(\mathit{i}\omega_n)$, we use a cubic spline to fit $\chi_{\alpha \beta}(\tau)$ and insert $10000$ sampling points on the imaginary time axis, and integrate the  oscillatory function as a function of $\tau$ using the composite trapezoidal rule.

For proxy M1, the errors from $\chi_{xx}(\mathit{i}\omega_n)$ are neglected and the error bar is $\pm 1$ standard error determined by jackknife resampling of $\chi_{xy}^{\text{ZF}}(\mathit{i}\omega_n)$. The cubic spline extrapolation to $\omega_n=0$ utilizes data on the first three non-zero $\omega_n$.
    The error bars for proxy M2 are constructed by error propagation of the standard errors in $\chi_{xy}^{\text{ZF}}(\mathit{i}\omega_1)$ and $\l[\chi_{xx}(\tau = \beta/2)\r]^2$, which are themselves determined by jackknife resampling.

\bibliography{mainNotes}

\end{document}